\documentclass{emulateapj}

\usepackage{color}
\usepackage{booktabs}
\usepackage{multirow}
\usepackage{graphicx}
\usepackage{threeparttable}

\renewcommand{\thefootnote}{\arabic{footnote}}
\begin{document}

\title{Large-Scale Filamentary Structures around the Virgo Cluster
  Revisited}

\author{Suk Kim\altaffilmark{1,2}, Soo-Chang Rey\altaffilmark{1,7}, 
Martin Bureau\altaffilmark{3}, Hyein Yoon\altaffilmark{4}, Aeree Chung\altaffilmark{4}, 
  Helmut Jerjen\altaffilmark{5}, Thorsten Lisker\altaffilmark{6}, 
  Hyunjin Jeong\altaffilmark{2}, Eon-Chang Sung\altaffilmark{2}, 
  Youngdae Lee\altaffilmark{1}, Woong Lee\altaffilmark{1}, and Jiwon Chung\altaffilmark{1}}

\altaffiltext{1}{Department of Astronomy and Space Science, Chungnam
  National University, 99 Daehak-ro, Daejeon 305-764, Korea; screy@cnu.ac.kr}
\altaffiltext{2}{Korea Astronomy \& Space Science institute, 776
  Daedeokdae-ro, Daejeon 305-348, Korea; star4citizen@kasi.re.kr}
\altaffiltext{3}{Sub-department of Astrophysics, Department of
  Physics, University of Oxford, Denys Wilkinson Building, Keble Road,
  Oxford OX1 3RH, UK}
\altaffiltext{4}{Department of Astronomy and Yonsei University
  Observatory, Yonsei University, Seoul 120-749, Korea}
\altaffiltext{5}{Research School of Astronomy and Astrophysics, The
  Australian National University, Cotter Road, Weston, ACT 2611,
  Australia}
\altaffiltext{6}{Astronomisches Rechen-Institut, Zentrum f\"ur
  Astronomie der Universit\"at Heidelberg (ZAH), M\"onchhofstra\ss e
  12-14, D-69120 Heidelberg, Germany}
\altaffiltext{7}{Author to whom any correspondence should be addressed}

\begin{abstract}
  We revisit the filamentary structures of galaxies around the Virgo
  cluster, exploiting a larger dataset based on the HyperLeda database
  than previous studies. In particular, this includes a large number
  of low-luminosity galaxies, resulting in better sampled individual
  structures. We confirm seven known structures in the distance range
  4~$h^{-1}$~Mpc~$<$ SGY~$<$ 16~$h^{-1}$~Mpc, now identified as
  filaments, where SGY is the axis of the supergalactic coordinate
  system roughly along the line of sight. The Hubble diagram of the
  filament galaxies suggests they are infalling toward the main-body
  of the Virgo cluster. We propose that the collinear distribution of
  giant elliptical galaxies along the fundamental axis of the Virgo
  cluster is smoothly connected to two of these filaments (Leo~II~A
  and B). Behind the Virgo cluster (16~$h^{-1}$~Mpc~$<$ SGY~$<$
  27~$h^{-1}$~Mpc), we also identify a new filament elongated toward
  the NGC~5353/4 group (``NGC~5353/4 filament'') and confirm a sheet
  that includes galaxies from the W and M clouds of the Virgo cluster
  (``W-M sheet''). In the Hubble diagram, the NGC~5353/4 filament
  galaxies show infall toward the NGC 5353/4 group, whereas the W-M
  sheet galaxies do not show hints of gravitational influence from the
  Virgo cluster. The filamentary structures identified can now be used
  to better understand the generic role of filaments in the build-up
  of galaxy clusters at z~$\approx$~0.
\end{abstract}

%\keywords{Filament, Sheet, Cluster of galaxies, Virgo cluster}

\section{Introduction}\label{sec:intro}

The visible matter distribution of the Universe forms a complex
web-like network composed of filaments and sheets separated by voids
(Bond et al.\ 1996; Arag{\'o}n-Calvo et al.\ 2010). Galaxies are
continuously funneled into higher density cluster environments through
these structures. In the hierarchical structure formation scenario,
z~$\approx$~0 filaments are the current end point of large-scale
structure evolution. Investigating filamentary structures around
dynamically young clusters should thus yield valuable information on
their assembly histories. Studying the large-scale galaxy distribution
around the Virgo cluster, the nearest rich, young cluster (Aguerri et
al.\ 2005), should therefore provide constraints on its formation and
evolution at the current epoch.

From three-dimensional mapping of nearby galaxies, Tully (1982,
hereafter T82) identified prolate and oblate over-densities of
galaxies around the Virgo cluster. These were however not clearly
revealed as conventional narrow filaments, but were rather sparse,
mainly due to the limited sample size. A better characterization of
these structures requires improved statistics from larger galaxy
samples, particularly those with fainter galaxies. 
% As the morphology-density relation implies that lower-density
% environments are populated by later-type galaxies ({\bf Dressler
% 1980}), probing the galaxy luminosity function to fainter limits is
% particularly important.

Recent spectroscopic surveys of galaxies such as the  Sloan Digital
Sky Survey (SDSS; e.g., SDSS Data Release 7, Abazajian et al.\ 2009)
do enable the exploration of possible filamentary
structures within an extensive volume around the Virgo
cluster. Moreover, radial velocity information on numerous galaxies
combined with distance data allows to investigate potential dynamical
relationships between the filaments and the Virgo cluster. The primary
goal of our study is thus to revisit the large-scale structures in the
vicinity of the Virgo cluster using an up-to-date statistically robust
dataset, to test whether the T82 structures can be better
characterized and identified as filaments. We also search for
previously unknown structures that may now be apparent. Finally, we
investigate whether the structures are physically connected to the
Virgo cluster using Hubble diagrams. Throughout this paper, we assume
a Hubble constant of $H_{0}$~= 100~$h$~km~s$^{-1}$~Mpc$^{-1}$, where
$h$~= 0.74 (Tully et al.\ 2008, hereafter T08).

\section{Data and analysis }
\label{sec:Data}

While the SDSS is large and homogeneous, it does not cover most of the
southern hemisphere and the spectroscopy is incomplete for both bright
($r$~$\lesssim$ 16) and faint ($r$~$\gtrsim$ 20) objects (e.g.,
Kim et al.\ 2014). We therefore adopt here the HyperLeda database
\footnote{http://leda.univ-lyon1.fr} (Paturel et al.\ 2003), which
while heterogeneous and necessarily incomplete does contain multiple
other surveys in addition to the SDSS Data Release 7 (Abazajian
et al.\ 2009).

For our analysis, we extracted all galaxies with available radial
velocities less than 6000~km~s$^{-1}$ in the region 115~deg~$<$
R.A.(J2000)~$<$ 240~deg and $-$35~deg~$<$ decl.(J2000)~$<$ 60~deg (see
Figure~1). This ensures the inclusion of all structures potentially
associated with the Virgo cluster (with a mean radial velocity of
$\approx$1000~km~s$^{-1}$ and a velocity dispersion of
$\approx$800~km~s$^{-1}$; Binggeli et al. 1993). However, as no
distinct structure in the vicinity of the Virgo cluster is identified
beyond a velocity of 3300~km~s$^{-1}$, we henceforth only discuss the
9168 galaxies within that range.
We examined the radial velocities of 755 SDSS filament galaxies, 
finding a mean difference of 6~km~s$^{-1}$ and a dispersion
of 63~km~s$^{-1}$ with respect to HyperLeda velocities (the latter
typically averaging multiple surveys). We therefore conclude that
the radial velocity information of our sample is not significantly
affected by the compilation of different sources in HyperLeda.
% {\bf We examined the radial velocities of 755 SDSS filament galaxies (see below), 
% finding a mean difference of 6~km~s$^{-1}$ and a dispersion
% of 63~km~s$^{-1}$ with respect to HyperLeda velocities (the latter
% typically averaging multiple surveys). We therefore conclude that
% the radial velocity information of our sample is not significantly
% affected by the compilation of different sources in HyperLeda.
% (MB: I am a bit worried by this comparison, since HyperLeda includes
% SDSS. The differences will therefore be biased low. A better
% comparison would be to exclude the SDSS data from HyperLeda first.)}

To construct an accurate spatial distribution of the galaxies, we
first converted the observed heliocentric radial velocities to
velocities relative to the centroid of the Local Group, and then
applied a correction for infall into the Virgo cluster (Mould et al.\
2000). Based on the assumption of a linear relationship between
redshift and distance, all galaxies were then mapped in the three
dimensions of the Cartesian supergalactic coordinate system: SGX, SGY,
SGZ (T08). Our search for filamentary structures was conducted within
the cuboid enclosed by the following limits:
\begin{center}
\noindent -25~$h^{-1}$~Mpc~$<$ SGX~$<$ 25~$h^{-1}$~Mpc\,\,\,, \\
\noindent \phantom{-2}4~$h^{-1}$~Mpc~$<$ SGY~$<$ 27~$h^{-1}$~Mpc\,\,\,, \\ 
\noindent -25~$h^{-1}$~Mpc~$<$ SGZ~$<$ 25~$h^{-1}$~Mpc\,\,\,. \\
\end{center}

The use of this extended region enables a more accurate
characterization of large-scale filamentary structures than the study
of T82. We rejected galaxies within a radius of 3.6~Mpc from the Virgo
cluster center (at least two times the virial radius; McLaughlin \ 1999), to
avoid contamination from cluster galaxies with large infall velocities
that are gravitationally bound to the cluster inside the infalling
region (Mamon et al.\ 2004; Falco et al.\ 2014). This leaves 8401
galaxies with which to search for large-scale structures.

The SGX-SGZ plane is best to look for large-scale structures, as
positional errors are small in this plane (roughly in the plane of the
sky) but substantial along the SGY axis (roughly along the line of
sight and thus affected by deviations from the Hubble flow). The
detection of filaments was thus performed by applying the following
steps:

\begin{enumerate}
\item We constructed a series of different SGX-SGY-SGZ volume slices
  with an arbitrary depth of 2~$h^{-1}$~Mpc along the SGY
  axis. Candidate structures were then selected by visual inspection
  of the SGX-SGZ projection of each slice, looking for overdense and
  long (i.e.,\ filamentary) galaxy distributions. If a candidate
  structure was continuously present in consecutive slices, we
  accordingly estimated its full range in three dimensions.

\item We performed three-dimensional third-order polynomial fitting
  with weighting by the local galaxy density to the visually selected
  candidate structures. If the standard deviation of the fit to a
  candidate structure was less than 1.5~$h^{-1}$~Mpc (comparable to or
  less than the typical thickness of filaments in simulations and
  observations; e.g.\ Colberg et al.\ 2005; Akahori \& Ryu 2010; Choi
  et al.\ 2010; Vazza et al.\ 2014), the structure was retained and
  classified as a filament. If not, the candidate structure was
  rejected. We further applied a two-sigma clip to the galaxies around
  the fitted lines (i.e.,\ the filament spines), to extract the
  galaxies that belong to each filament.

\item The authenticity of each candidate structure was verified by
  looking at a multitude of SGX-SGY-SGZ projections, and the more
  diffuse (i.e.,\ non-filamentary) nature of the rejected structures
  was assessed (chance projections, sheets, etc).
\end{enumerate}

With the search strategy above, we identified seven filaments and one
sheet in the volume surrounding the Virgo cluster, divided into two
subsamples with distinct SGY ranges (see Sec.\ 3). Figure~1 shows the
spatial distribution of the 1013 galaxies belonging to these
structures in the equatorial coordinate system. The properties of
these structures are summarized in Table~1. We note that all
structures mostly consist of faint galaxies (M$_{B}>-$19;
$\approx$88$\%$ of the total sample).

Finally, we collected redshift-independent distances of sample
galaxies from the NASA/IPAC Extragalactic
Database\footnote{https://ned.ipac.caltech.edu} (NED), with 229
matches. We calculated the distances ($R_{\rm VC}$) and radial
velocities ($V_{\rm VC}$) of these galaxies relative to the Virgo
cluster center (Karachentsev \& Nasonova 2010), adopting a distance of
the Virgo cluster from us of 16.5~Mpc (Mei et al.\ 2007). 
In this instance, for the calculation of $V_{\rm VC}$
(T08), the heliocentric radial velocities were corrected for the
motion of the local sheet and local void only (and {\em not} for
Virgo-centric infall). Finally, peculiar velocities ($PV_{\rm VC}$)
were derived using the following equation:
\begin{equation}
PV_{\rm VC} = V_{\rm VC} - R_{\rm VC} \times H_{0}\,\,\,.
\end{equation}

\section{Result} \label{sec:Result}

\subsection{Virgo-related structures $\rm ($4$\,\,h^{-1}$~$\rm Mpc~< SGY~<$
  16$\,\,h^{-1}$~$\rm Mpc)$}

Figure~2 presents the distribution of six filaments with
4~$h^{-1}$~Mpc~$<$ SGY~$<$ 16~$h^{-1}$~Mpc in the SGX-SGZ projected
plane (Fig.~2a) and the SGX-SGY-SGZ three-dimensional space
(Fig.~2b). We named these filaments after the structures designated as
clouds or spurs by T82 (Fig.~2c). By construction, all the filaments
are narrower and longer than the broader and more diffuse T82
structures. All are also elongated toward the Virgo cluster,
suggesting they are related to it.

It is interesting to note that when we apply our search strategy to
the T82 sample, the T82 structures are {\em not} identified as
filaments. This is most likely due to the small T82 sample size, as
the number of galaxies in each filament is two or three times larger
in our sample.

In addition, while the majority of galaxies in T82 are bright
(M${_B}<-$19), our structures are mainly composed of less luminous
galaxies (M${_B}>-$19). Bright galaxies (large open circles in
Figs.~1\,--\,4) exhibit an uneven distribution with numerous gaps
along the filaments. Conversely, the large number of low-luminosity
galaxies available here (small filled circles in Figs.~1\,--\,4) helps
to better delineate continuous filaments by filling in the gaps.

The Leo~II cloud appears to be a multi-stem clump rather than a single
prolate one (Fig.~2c). Indeed, in a more extensive area around the
Leo~II cloud, we identified two filaments, Leo~II~A and B. The
Leo~II~B filament traces a sparse region of the cloud, whereas the
Leo~II~A filament includes the main clump of the Leo~II cloud and is
one of the largest structures we detect (16~$h^{-1}$~Mpc long in the
SGX-SGZ plane), with a curved shape. It is known that warped or 
irregular filaments are more common and are on average longer than 
straight filaments (Colberg et al.\ 2005).

Over 150 galaxies in the six filaments identified in Figs.~2a and 2b
have a redshift-independent distance. Their Hubble diagram in the
Virgo-centric reference frame is presented in Figure~2d. Most galaxies
in all these filaments exhibit a distinct deviation of their radial
velocities from an unperturbed Hubble flow (red line; see also the
median values in the inset), suggesting appreciable infall toward the
Virgo cluster. We also plot a model of the radial infall velocity
profile of the Virgo cluster (blue line), adopting the model of Falco
et al.\ (2014; eq.~6, representing the mean velocity of infalling
galaxies in the outer regions of the cluster) with a virial radius of
1.55~Mpc (McLaughlin\ 1999) and a virial velocity of
800~km~s$^{-1}$ (Binggeli et al.\ 1993). The observed distribution of
radial velocities of the filament galaxies is entirely consistent with
the expected infall profile, confirming that all these filaments are
dynamically connected to the Virgo cluster (see also Karachentsev et
al.\ 2014).

\subsection{Structures behind the Virgo cluster
  $\rm ($16$\,\,h^{-1}$~$\rm Mpc~< SGY~<$ 27$\,\,h^{-1}$~$\rm Mpc)$}

Figures~3a and 3b are analogous to Figs.~2a and 2b but for
16~$h^{-1}$~Mpc~$<$ SGY~$<$ 27~$h^{-1}$~Mpc. They show one filament and
one sheet identified behind the Virgo cluster, at mean distances of 33
and 41~Mpc from us, respectively.

The filament (red circles), which we call the NGC~5353/4 filament, is
long and thin and extends out from the NGC~5353/4 group, running
tangentially past the Virgo cluster rather than pointing toward
it. Filaments running from the vicinity of the NGC~5353/4 group to the
Virgo (e.g., Canes Venatici filament, see the gray rectangle in Fig.~3a)
 and Coma clusters in different SGY ranges have been reported in
the past (Tully $\&$ Trentham 2008; Pomar{\`e}de et al.\ 2015), but the
NGC5353/4~filament identified here is not specifically referred to in
previous studies.

Seventeen galaxies in the NGC~5353/4 filament have a
redshift-independent distance. Their Hubble diagram in the NGC~5353/4
group-centric reference frame is presented in Fig.~3c. All but one
NGC~5353/4 filament galaxies show a clear negative offset from the
Hubble flow, with a mean of $-$242~km~s$^{-1}$, indicating that they
are infalling toward the NGC~5353/4 group. Recently, Pomar{\`e}de et
al.\ (2015) identified the Arrowhead mini-supercluster, in which the
NGC~5353/4 group is located, at the edge of the Arrowhead flow pattern
in the velocity field. They also suggested that the NGC~5353/4 group
is located at the boundary of the flow pattern of the Laniakea
supercluster (Tully et al.\ 2014). The NGC~5353/4 filament thus
appears to be a small structure bridging the Arrowhead and Laniakea
superclusters (see arrows in Fig.~3a for the directions to the two
superclusters).

A thin sheet structure (yellow circles) is also clearly visible in
Figs.~3a and 3b. While it appears filament-like in the projected plane
(Fig.~3a), it is clearly flattened (Fig.~3b), with an axial ratio
SGX:SGY:SGZ = 9:6:1. Since this structure includes the W and M clouds
as well as the western part of the southern extension of the Virgo
cluster, we name it the W-M sheet. The existence of this sheet was
however alluded to by Binggeli et al.\ (1993), and a connection
between the W and M clouds was also suggested (e.g.\ Ftaclas et al.\
1984; Yoon et al.\ 2012).

Fifty four galaxies in the W-M sheet have a redshift-independent
distance. Their Hubble diagram in the Virgo-centric reference frame is
presented in Fig.~3d. The W-M sheet galaxies have a large scatter
around the Hubble flow with no trend, indicating that the sheet is
unlikely to be dynamically connected to the Virgo cluster.

\section{DISCUSSION AND CONCLUSIONS}
\label{sec:DISCUSSION}

Cosmological numerical simulations suggest that galaxy clusters have
non-spherical shapes, owing to the preferential infall of matter
along the “cosmic web” (e.g.\ Limousin et al.\ 2013 and references
therein). Dynamically young clusters within filaments are also
expected to be more elongated than relaxed clusters.
Moreover, it has been noted that the Virgo cluster exhibits a noticeably
non-spherical shape, elongated along a direction almost parallel to
the line of sight, albeit primarily when considering galaxies near the
cluster center (e.g.\ Binggeli et al.\ 1987; West $\&$ Blakeslee 2000;
Mei et al.\ 2007). 

In Figure~4a, we show the spatial distribution of
galaxies in the Virgo cluster using only galaxies classified as
``certain'' cluster members in the Extended Virgo Cluster Catalog
(EVCC, Kim et al. 2014; black open circles inside a large rectangular
box). Using HyperLeda, we also plot the galaxies (gray open circles)
within the cluster zero-velocity surface (large gray circle with a
26~deg radius; Karachentsev $\&$ Nasonova 2010). Noticeably, the
bright elliptical galaxies (M$_{B}<-$19; red open circles) are all
aligned. The galaxies in the western outskirts of the cluster (i.e.,\
180~deg~$<$ R.A.~$<$ 160~deg) also exhibit a filamentary distribution,
with a major axis similar to that of the bright ellipticals nearer the
cluster center. Moreover, this collinear distribution of cluster
galaxies appears to be smoothly connected to the Leo~II~A and B
filaments identified here (cyan and green circles).

The genuine alignment of the galaxies in three dimensions is confirmed
by the distribution of the galaxies with available
redshift-independent distances in the plane of R.A.\ vs.\ distance
from us (Fig.~4b). Instead of a random distribution, the galaxy
distances are systematically increasing from the cluster center to the
filaments. This systematic trend supports the suggestion that the
galaxy distribution along the fundamental axis of the Virgo cluster is
associated with the Leo~II~A and B filaments. West $\&$ Blakeslee
(2000) claimed that the collinear distribution of the brightest
ellipticals near the cluster centre may be part of a filament
connecting the Virgo cluster to Abell~1367 (see the arrow indicating
the direction to Abell~1367 in Fig.~4b), but our data suggest that a
connection to the Leo~II~A and B filaments is more likely. In any
case, our results are clear observational evidence that the spatial
distribution of cluster galaxies is not completely spherically
symmetric, reflecting the anisotropy of their accretions along
filaments (e.g.\ Joachimi et al.\ 2015 and references therein; see
also Lee et al.\ 2014). In this respect, comparative studies of
galaxies in Virgo-related filaments with ones in the cluster itself
should help to understand the growth and evolution of the cluster (S.\
Kim, in preparation).

Filament galaxies are a crucial tool to measure galaxy cluster and
group parameters. We have already estimated a new Virgo cluster
dynamical mass by applying the novel method of Falco et al.\ (2014),
based on the universal radial velocity profile of filament galaxies,
to the Leo~II~A and Virgo~III filaments (Lee et al.\ 2015a). We have
also determined the turn-around radius of the NGC~5353/4~group by
applying the Falco et al.\ methodology to the NGC 5353/4~filament (Lee
et al.\ 2015b). Accurate measurements of the physical parameters of
nearby galaxy clusters and groups using filaments in their vicinity
can also provide independent constraints on cosmological models (see
Lee et al.\ 2015a,b for details).

Finally, filaments are prime targets for detailed investigations of
the physical processes controlling the transition of field galaxies to
cluster galaxies (e.g.\ Balogh et al.\ 2004; Ebeling et al.\ 2004;
Porter et al.\ 2008; Yoon et al.\ 2012; Guo et al.\ 2015;
Mart{\'{\i}}nez et al.\ 2016). Indeed, as the majority of filament
galaxies are faint dwarf galaxies, with low binding energies, they are
easily affected by even weak perturbations. They are thus well-suited
to probing the details of the multiple mechanisms (e.g.\ gas stripping
and tidal interactions) that can affect galaxies in low-density
environments before they enter higher-density regions
(``pre-processing''; Fujita 2004; Cybulski et al.\ 2014). To directly
probe environmental effects in filaments, we are thus currently
examining the H\,{\small I} morphology and kinematics of carefully
selected late-type galaxies in filaments around the Virgo cluster
(e.g.\ Yoon et al.\ 2015; H.\ Yoon, in preparation). 
In a forthcoming paper, we will also compare dwarf galaxies in
filaments with those in other environments, including cluster
(e.g.\ Kim et al.\ 2014), group (e.g.\ Pak et al.\ 2014), and the
field (S.\ Kim, in preparation).

\begin{acknowledgments}
We thank the anonymous referee for the clarifications and comments
that helped to improve the original manuscript.
We thank D.S.\ Ryu and J.H.\ Lee for helpful discussions. 
This research was supported by the Basic Science Research Program through the National Research Foundation of Korea (NRF) funded by the Ministry of Education, Science, and Technology (2015R1A2A2A01006828). 
Support was also provided by the NRF of Korea to the Center for Galaxy Evolution Research (No.\ 2010-0027910). 
S.K.\ acknowledges support from a National Junior Research Fellowship of NRF (No.\ 2011-0012618). 
H.J.\ acknowledges support from the NRF funded by the Ministry of Education (NRF-2013R1A6A3A04064993).
This study was also financially supported by the research fund of Chungnam National University in 2014.
This work has been also supported by Science Fellowship of POSCO TJ Park Foundation and the NRF grant No. 2015R1D1A1A01060516.
HJ acknowledges the support of the Australian Research Council through Discovery Project DP150100862.
  
\end{acknowledgments}

\newpage
%=======TABLE=======
\begin{table*}
\tabletypesize{\scriptsize} 
%\caption{Filaments (and Sheets) around the Virgo Cluster}
\label{my-label}
\resizebox{\textwidth}{!}{\hskip-0.0cm\begin{tabular}{ccccccccccccc}
%\begin{tabular}{lcccccccccccc}
\tablewidth{0pt}
\hline
\hline
     &                &                &                &               &              &                        &                        &                  & \multicolumn{2}{c}{This work}        & \multicolumn{2}{c}{Tully (1982)} \\ \cline{10-13} 
Name & SGX            & SGY            & SGZ            & cz
                                                                        &Length        & $R_{\rm VC}$               & Peculiar velocity      & Distance$_{_{\rm MW}}$  & N & N$_{\rm faint}$                      & N    &     N$_{\rm faint}$          \\ 
     & ($h^{-1}$ Mpc) & ($h^{-1}$ Mpc) & ($h^{-1}$ Mpc) & (km s$^{-1}$) &($h^{-1}$ Mpc)& (Mpc)                  & (km s$^{-1}$)          & (Mpc)            &   &                                  &      &                            \\

\hline
(1)          & (2)                 & (3)                & (4)                 & (5)              & (6)  & (7)                & (8)                      & (9)                  & (10)                  & (11)                  & (12)                       & (13)         \\ 
\hline 
Leo II A     &  \phantom{-1}0.21  $\sim$ \phantom{-}10.36 & \phantom{1}9.26  $\sim$ 15.05 & -15.47 $\sim$ \phantom{1}-4.16 & 1171 $\sim$ 2267 & 16.0 & 11.71 $\sim$ 46.68 & -213.98\phantom{$^{*}$} (283.02)   & 26.30 \phantom{0}(4.86)   &180  &165   & \multirow{2}{*}{97$^{**}$} & \multirow{2}{*}{45$^{**}$} \\
Leo II B     &  \phantom{-1}0.30  $\sim$ \phantom{-}15.65 & 10.90 $\sim$ 14.56 & \phantom{1}-9.88  $\sim$ \phantom{1}-3.51 & 1257 $\sim$ 2267 & 15.5 & 10.84 $\sim$ 32.19 & -282.04\phantom{$^{*}$} (306.53)         & 26.40 \phantom{0}(4.71)         & 105                   & \phantom{1}94                    &                            &                             \\
Leo Minor    &  \phantom{-1}0.55  $\sim$  \phantom{-1}5.89 & \phantom{1}4.11  $\sim$  \phantom{1}6.49 & \phantom{1}-2.61  $\sim$ \phantom{1}-0.99 & \phantom{1}505  $\sim$ \phantom{1}772  & \phantom{1}5.4  &  \phantom{1}7.27 $\sim$ 17.41 & -250.09\phantom{$^{*}$} (137.90)         & 14.07 \phantom{0}(3.65)         & \phantom{1}54   & \phantom{1}48    & 46\phantom{$^{**}$}                         & 31\phantom{$^{**}$}                  \\
Canes Venatici &  \phantom{-1}0.78  $\sim$  \phantom{-1}4.37 & \phantom{1}6.88  $\sim$ 13.92 & \phantom{-1}1.38   $\sim$ \phantom{-1}4.80  & \phantom{1}674  $\sim$ 1446 & \phantom{1}4.8  &  \phantom{1}7.15 $\sim$ 27.05 & -254.95\phantom{$^{*}$} (356.27)         & 20.96 \phantom{0}(6.83)         & \phantom{1}51   & \phantom{1}48    & 18\phantom{$^{**}$}                         & 14\phantom{$^{**}$}                   \\
Virgo III    & -10.56 $\sim$ \phantom{1}-3.91 & \phantom{1}9.50  $\sim$ 15.57 & \phantom{-1}2.35   $\sim$ \phantom{-}11.72 & 1160 $\sim$ 2196 & 11.4 &  \phantom{1}7.26 $\sim$ 32.07 & -102.24\phantom{$^{*}$} (364.43)         & 26.70 \phantom{0}(6.00)         & 181  & 162   & 61\phantom{$^{**}$}                         & 20\phantom{$^{**}$}                  \\
Crater       & -12.25 $\sim$ \phantom{1}-4.62 & \phantom{1}8.36  $\sim$ 12.70 & \phantom{1}-5.91  $\sim$ \phantom{1}-2.98 & 1436 $\sim$ 1903 & \phantom{1}7.9  &  \phantom{1}8.28 $\sim$ 19.13 &-148.95\phantom{$^{*}$} (138.52)          & 23.31 \phantom{0}(3.55)         & \phantom{1}84   & \phantom{1}69    & 35\phantom{$^{**}$}                         & 6\phantom{$^{**}$}                   \\
NGC5353/4    & -16.04 $\sim$  \phantom{-1}4.23 & 21.71 $\sim$ 26.53 & \phantom{1}-1.19  $\sim$ \phantom{-1}8.92  & 2268 $\sim$ 3238 & 21.9 &  \phantom{1}6.70 $\sim$ 32.27 &-242.46$^{*}$ (306.19)    & 41.05 \phantom{0}(7.79)         & 102  & \phantom{1}89    & -\phantom{$^{**}$}                          & -\phantom{$^{**}$}                   \\
W-M sheet    & -13.38 $\sim$ \phantom{1}-1.66 & 16.03 $\sim$ 24.99 & \phantom{1}-3.10  $\sim$ \phantom{1}-1.10 & 1806 $\sim$ 2968 & 11.9 &  \phantom{1}1.45 $\sim$ 65.64 &-108.85\phantom{$^{*}$} (786.05)          & 32.60 (10.91)        & 256  & 221   & -\phantom{$^{**}$}                         & -\phantom{$^{**}$}                   \\ 
\hline 
\end{tabular}
}
\begin{tablenotes}
\scriptsize
\item $^{*}$Median peculiar velocity and standard deviation in the
  NGC5353/4 group-centric reference frame.
\item $^{**}$Tully (1982) designated the Leo~II cloud as a single
  structure.
\end{tablenotes}
\begin{tablenotes}
\item Note for columns. (1) Name of the structure. (2)--(4) Range of
  the structure in supergalactic coordinates. (5) Range of the
  structure in radial velocity. (6) Length of the structure in the
  SGX-SGZ plane. (7) Range of the distances relative to the Virgo
  cluster center. (8) Median value and standard deviation of the
  peculiar velocities in the Virgo-centric reference frame. (9) Median
  value and standard deviation of the distances from the Milky
  Way. (10)--(11) Number of total and faint (M$_{B}>-$19)
  galaxies. (12)--(13) Number of total and faint (M$_{B}>-$19)
  galaxies in Tully (1982).
\end{tablenotes}
\end{table*}
%===============

\begin{figure*}
  \figurenum{1}
  \plotone{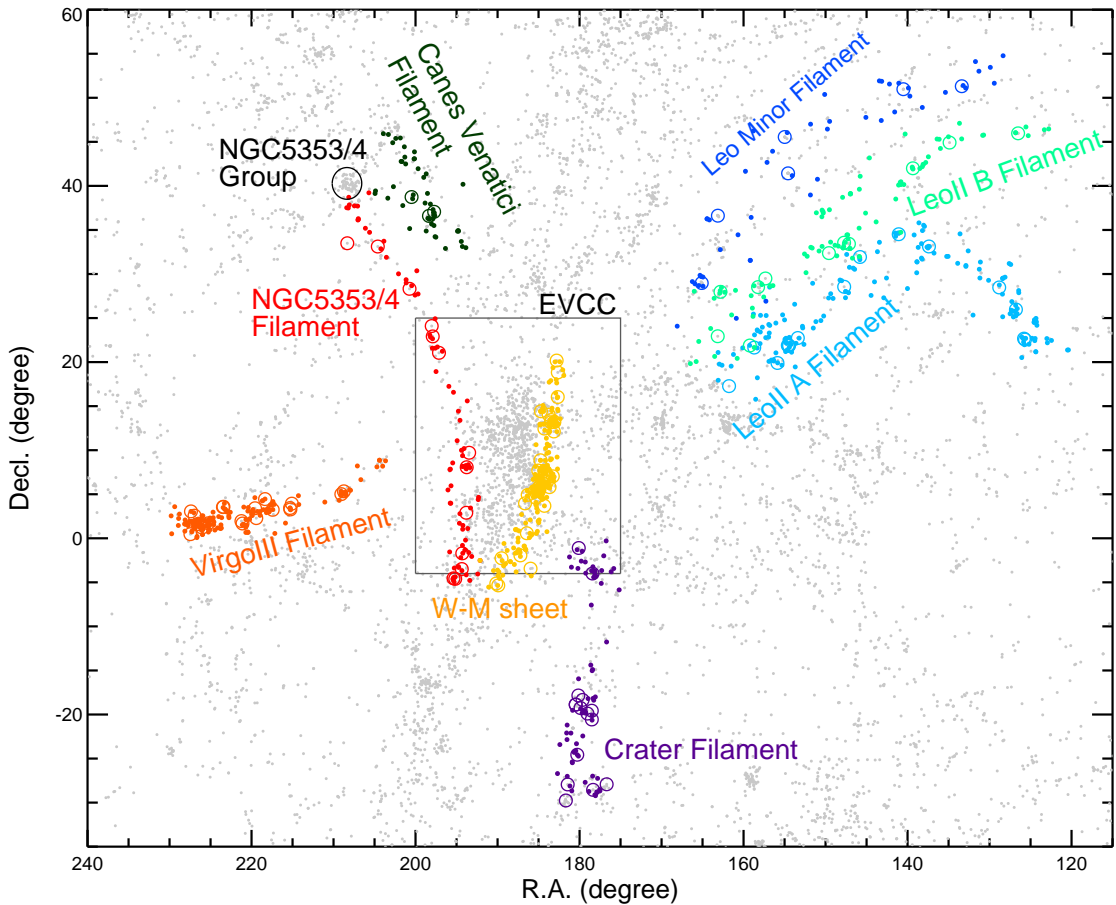}
  \caption{Spatial distribution of galaxies in seven filaments and one
    sheet around the Virgo cluster in the equatorial coordinate
    system. Bright (M$_{B}<-$19) and faint (M$_{B}>-$19) galaxies are
    denoted by large open circles and small filled circles,
    respectively. Different colors denote different structures. The
    large rectangular box is the region of the Extended Virgo Cluster
    Catalog (EVCC; Kim et al.\ 2014). Gray dots represent galaxies not
    associated with the Virgo cluster or a particular structure.}
  \label{fig:fig1}
\end{figure*}

\begin{figure*}
  \figurenum{2}
  \plotone{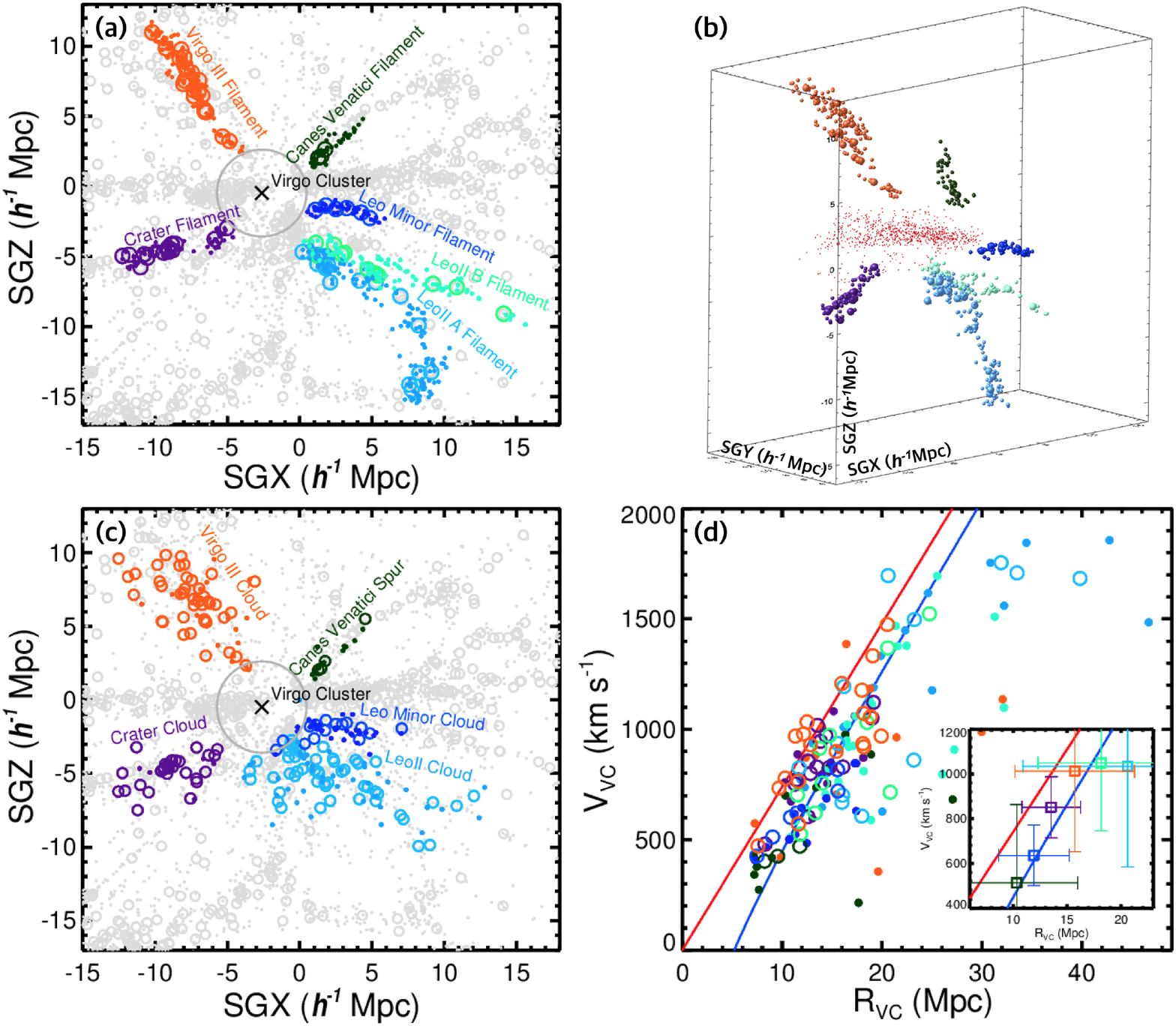}
  \caption{Spatial distribution (a--c) and Hubble diagram (d) of six
    filaments in the range 4~$h^{-1}$~Mpc~$<$ SGY~$<$
    16~$h^{-1}$~Mpc. Symbols are the same as in Fig.~1. (a) Projected
    spatial distribution of the filaments in the SGX-SGZ plane. The
    large gray circle marks two virial radii around the Virgo
    cluster. (b) Three-dimensional distribution of the filaments. Red
    dots are Virgo cluster galaxies in the EVCC. (c) Same as (a) for
    the structures mentioned by Tully (1982). (d) Hubble diagram of
    the filament galaxies in the Virgo-centric reference frame. The
    red and blue lines indicate the Hubble flow and a model of the
    radial infall velocity profile caused by the gravitational pull of
    the Virgo cluster, respectively. The inset shows the median
    Virgo-centric radial velocity and distance of each filament (error
    bars indicate one standard deviation).}
  \label{fig:fig2}
\end{figure*}

\begin{figure*}
  \figurenum{3}
  \plotone{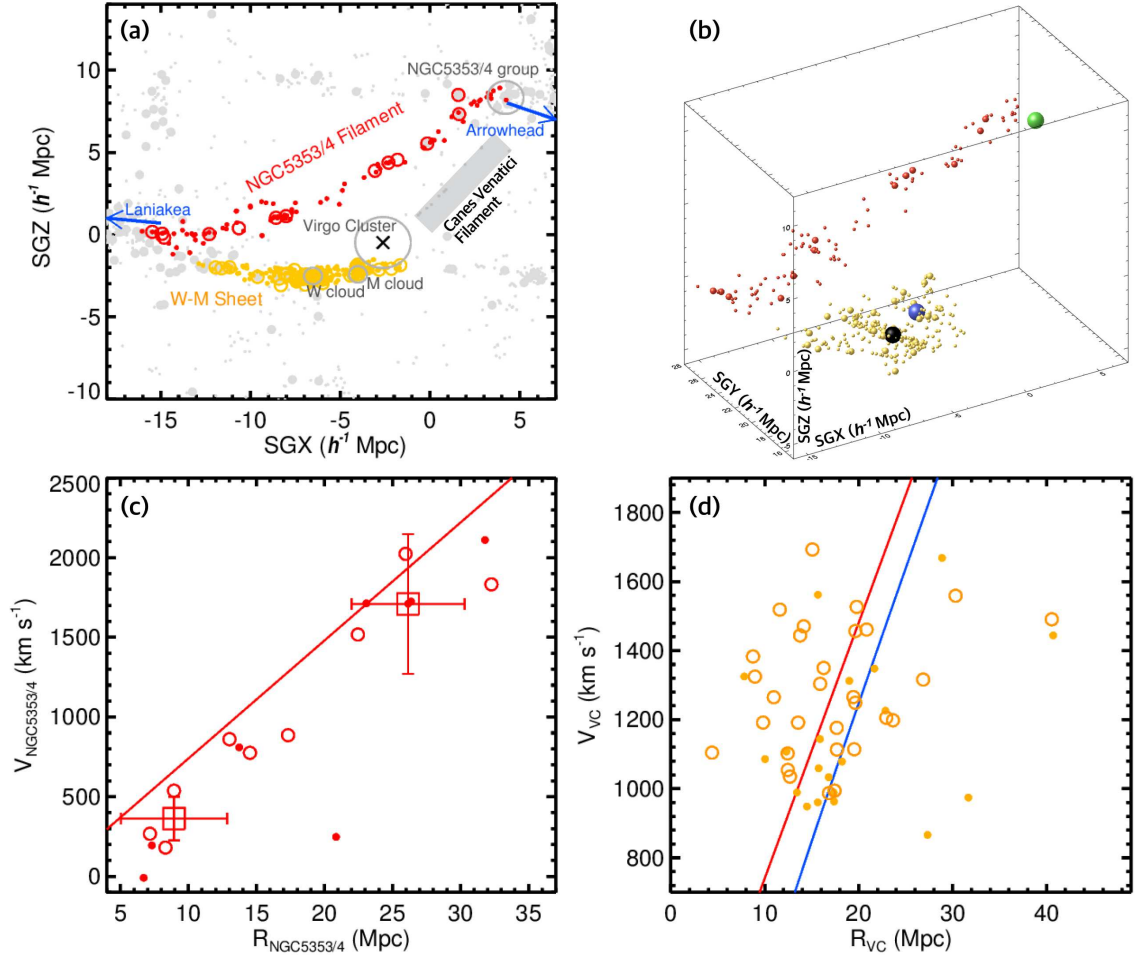}
  \caption{Spatial distribution (a--b) and Hubble diagram (c--d) of
    the NGC~5353/4 filament (red circles) and the W-M sheet (yellow
    circles) in the range 16~$h^{-1}$~Mpc~$<$ SGY~$<$ 27~$h^{-1}$
    Mpc. Symbols are the same as in Fig.~1. (a) Projected spatial
    distribution of the structures in the SGX-SGZ plane. The gray
    rectangle marks the region of the Canes Venatici filament. The
    directions to the Arrowhead and Laniakea superclusters are
    indicated with blue arrows (Tully et al.\ 2014; Pomar{\`e}de et
    al.\ 2015). (b) Three-dimensional distribution of the
    structures. The black, blue, and green spheres denote the W cloud,
    M cloud, and NGC~5353/4 group, respectively. (c) Hubble diagram of
    the NGC~5353/4 filament galaxies in the NGC~5353/4 group-centric
    reference frame. The red line indicates the Hubble flow, while the
    open squares denote the median NGC~5353/4 group-centric radial
    velocity and distance of galaxies in different distance ranges
    (error bars indicate one standard deviation). (d) Hubble diagram
    of the W-M sheet galaxies in the Virgo-centric reference
    frame. The red and blue lines indicate the Hubble flow and a model
    of the radial infall velocity profile caused by the gravitational
    pull of the Virgo cluster, respectively.}
  \label{fig:fig3}
\end{figure*}

\begin{figure*}
  \figurenum{4}
  \plotone{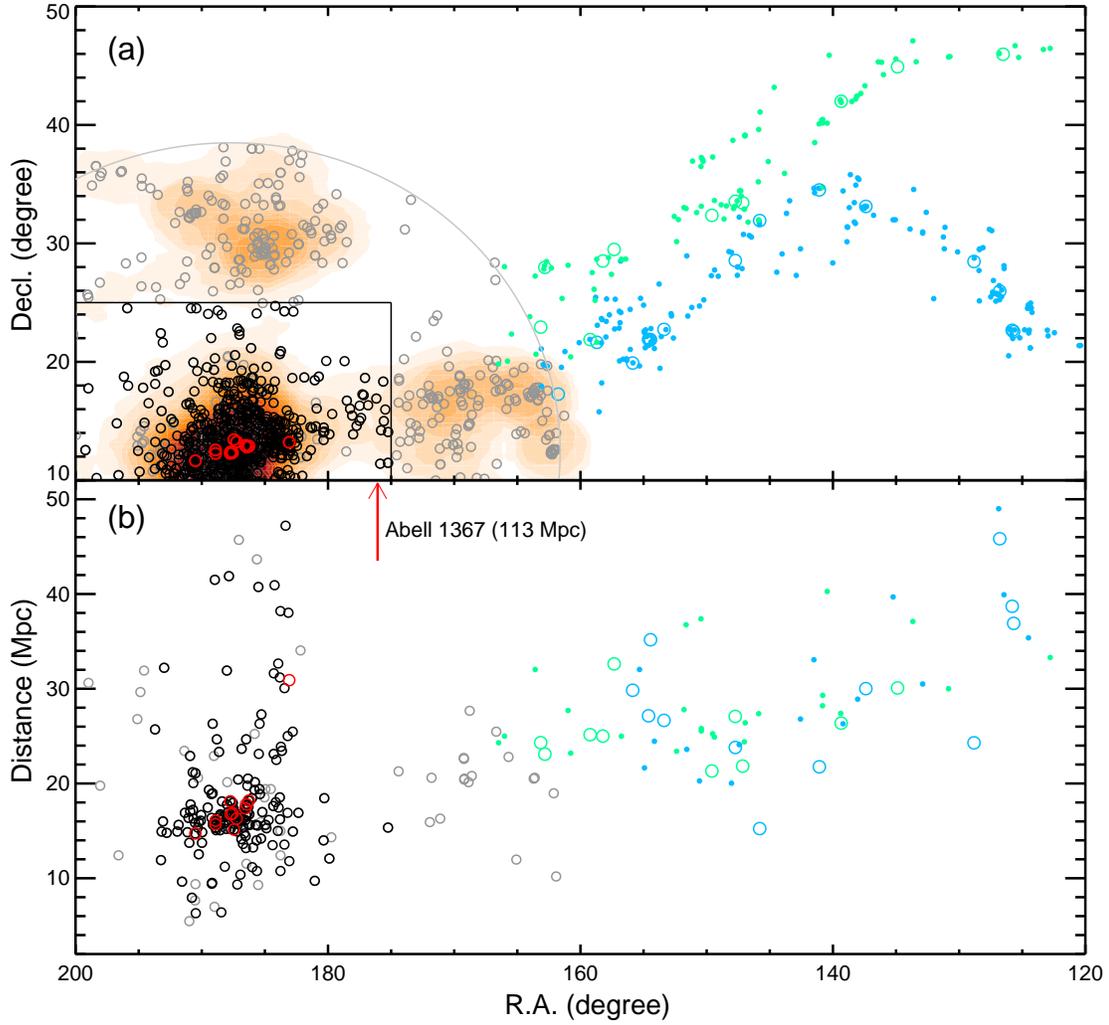}
  \caption{(a) Spatial distribution of galaxies in the Leo~II~A (cyan
    circles) and B (green circles) filaments. Red and black open
    circles show bright elliptical galaxies (M$_{B}<-$19) and
    ``certain'' Virgo cluster member galaxies in the region of the
    EVCC (large rectangular box), respectively. Gray open circles show
    galaxies in the range 4~$h^{-1}$~Mpc~$<$ SGY~$<$ 16~$h^{-1}$ Mpc
    from HyperLeda and orange/red contours denote the local number density distribution
		of these galaxies. The large gray circle is the upper limit of the
    zero-velocity surface (26~deg radius). (b) Same as (a) but in the
    R.A.\ vs.\ distance from us plane. The direction to the Abell~1367
    cluster and its distance from us are indicated by the annotated
    red arrow.}
  \label{fig:fig4}
\end{figure*}

\end{document}